\documentstyle[twoside,titlepage,psfig,12pt]{article}
\newcommand{\gtsima}{$\; \buildrel > \over \sim \;$}
\newcommand{\ltsima}{$\; \buildrel < \over \sim \;$}
\newcommand{\simgt}{\lower.5ex\hbox{\gtsima}}
\newcommand{\simlt}{\lower.5ex\hbox{\ltsima}}

\newcommand{\T}{ {\scriptscriptstyle {\rm T}} }
\newcommand{\sz}{ {{\rm sz}} }
\newcommand{\COBE}{ {\scriptscriptstyle {\rm COBE}} }

\newcommand{\keV}{ {\rm keV} }
\newcommand{\mpc}{ {\rm Mpc} }
\newcommand{\dttsz}{{\hbox
      {$\displaystyle\left({\delta T \over T}\right)_\sz$} }}
\newcommand{\dtt}{{\hbox
      {$\displaystyle\left({\delta T \over T}\right)$} }}
\def\pp{\par\parshape 2 0truecm 15.5truecm 1truecm 14.5truecm\noindent}

\begin{document}
\begin{titlepage}

\addtocounter{footnote}{0}

\vspace*{2cm} 
\begin{center}

\baselineskip24pt

{\bf  
THE X-RAY HALO OF THE LOCAL GROUP\\
AND ITS IMPLICATIONS\\
FOR MICROWAVE AND SOFT X-RAY BACKGROUNDS\\

}

\end{center}
\vspace{2.0cm}
 
{\sc Yasushi Suto$^{1,2}$, 
Kazuo Makishima$^{1,2}$, 
Yoshitaka Ishisaki$^1$, \& Yasushi Ogasaka$^3$}

\begin{center}

\vspace{0.5cm}

\noindent $^1$ {\it Department of Physics, The University of Tokyo,
Tokyo 113, Japan}

\noindent $^2$ {\it RESCEU (Research Center for the Early Universe),
School of Science,\\
The University of Tokyo, Tokyo 113, Japan}

\noindent $^3$ {\it Institute of Space and Astronautical Science,
3-3-1 Yoshinodai, Sagamihara, \\ Kanagawa 229, Japan}

e-mail: suto@phys.s.u-tokyo.ac.jp, maxima@phys.s.u-tokyo.ac.jp, 
    ishisaki@miranda.phys.s.u-tokyo.ac.jp, ogasaka@astro.isas.ac.jp,

\vspace{0.5cm}

Received  1995 December 26; accepted 1996 January 31

\end{center}

\vfill
\centerline {\it Astrophysical Journal (Letters), in press.}

\end{titlepage}

\newpage

\vspace{0.5cm}

\centerline {\bf ABSTRACT}
\baselineskip=14pt
\baselineskip=24pt

Since recent X-ray observations have revealed that most clusters of
galaxies are surrounded by an X-ray emitting gaseous halo, it is
reasonable to expect that the Local Group of galaxies has its own
X-ray halo. We show that such a halo, with temperature $\sim 1\keV$
and column density $\sim O(10^{21}) {\rm cm}^{-2}$, is a possible
source for the excess low-energy component in the X-ray background.
The halo should also generate temperature anisotropies in the
microwave background via the Sunyaev-Zel'dovich effect. Assuming an
isothermal spherical halo with the above temperature and density, the
amplitude of the induced quadrupole turns out to be comparable to the
COBE data without violating the upper limit on the $y$-parameter.  The
induced dipole is negligible compared to the peculiar velocity of the
Local Group, and multipoles higher than quadrupole are generally much
smaller than the observed ones. However non-sphericity and/or
clumpiness of the halo will produce a stronger effect.  Therefore the
gaseous halo of the Local Group, if it exists, will affect the
estimate of the primordial spectral index $n$ and the amplitude of the
density fluctuations deduced from the {\sl COBE} data.
\medskip

\pp {\it Subject headings}: cosmic microwave background ---
cosmology: theory --- diffuse radiation --- Local Group\\

\newpage
\setcounter{page}{2}

\begin{center}
\section{\normalsize\bf INTRODUCTION}
\end{center}
\setcounter{equation}{0}

\baselineskip=24pt

The origin of the X-ray background (XRB) remains one of the most
challenging problems in X-ray astrophysics. Figure 1 summarizes the
current results on the XRB energy spectra, $I(\varepsilon)$ below
10\keV observed with different satellites.  As is known (McCammon \&
Sanders 1990; Fabian \& Barcons 1992), $I(\varepsilon) \sim 10
\varepsilon^{-0.4}{\rm keV}\cdot{\rm s}^{-1}\cdot{\rm
sr}^{-1}\cdot{\keV}^{-1}$ is a good empirical fit over the range 3 to
10 \keV, where $\varepsilon$ denotes X-ray energy in units of
\keV. Both Einstein IPC (Wu et al. 1991; plotted in diamonds) and
Rosat (Hasinger 1992; Shanks et al. 1991; upper-left lines) data
suggest a large excess below 2\keV. More recently the Japanese X-ray
satellite ASCA (crosses) reported a modest but significant excess soft
component below 1\keV relative to the extrapolation of the above
power-law fit in the higher energy band (Gendreau et al. 1995). Thus
the existence of the soft excess is well established although its
amplitude is still somewhat controversial.

Since Galactic absorption becomes important in the soft X-ray energy
band, the origin of this excess component allows several possibilities
including Galactic sources, extragalactic point-like sources (Hasinger
1992; Shanks et al. 1991), a diffuse thermal component (Wang \& MaCray
1993), the accumulation of the thermal bremsstrahlung emission from
distant clusters of galaxies (Kitayama \& Suto 1996).  Cen et
al. (1995) found an excess of soft X-ray background below 1 \keV from
their hydrodynamical simulations which properly incorporate the line
emissions as well as the thermal bremsstrahlung emission; the excess
is mainly originated from the the low temperature and low density
plasma surrounding distant clusters of galaxies since cooler
background gas produces much stronger line emissions.  Yet another
possibility which we propose here is the emission from an X-ray halo
of the Local Group (LG). Since the gaseous halos of clusters of
galaxies are known to be strong sources of X-rays, it is reasonable to
assume that the LG has its own X-ray halo.

\newpage

\begin{center}
\section{\normalsize\bf SUNYAEV-ZEL'DOVICH EFFECT DUE TO THE HALO OF
THE LOCAL GROUP}
\end{center}
\setcounter{equation}{0}

To be more specific, suppose that the LG is associated with a
spherical isothermal plasma whose electron number density is given by
\begin{equation}
\label{eq:ner}
n_e(r) = n_0 {r_c^2 \over r^2 + r_c^2},
\end{equation}
where $n_0$ is the central density and $r_c$ is the core radius.  If
we are located at distance $x_0$ off the LG center
(Fig. \ref{fig:lghalo}), the electron column density at angular
separation $\theta$ from the direction to the center is
\begin{equation}
\label{eq:ne}
N_e(\mu) = \int_0^{\infty} { n_0 r_c^2 d\xi \over 
\xi^2 -2x_0\mu \xi + x_0^2 + r_c^2 } 
= {n_0 r_c^2 \over x_0}{1 \over \sqrt{a^2 - \mu^2}}
\left[{\pi \over 2}+\sin^{-1}\left({\mu \over a}\right)\right] ,
\end{equation}
where $\mu \equiv \cos\theta$, and $a\equiv \sqrt{1+(r_c/x_0)^2}$.  In
Fig.1 the LG halo contribution to the XRB spectra simulated with the
Raymond-Smith model assuming an isothermal plasma temperature
$T=1\keV$, $r_c=0.15\mpc$, $x_0=0$, $n_0 = 10^{-4} {\rm cm}^{-3}$ and
0.3 times solar abundances is also plotted(lower-left histogram).  If
we add this component to $I(\varepsilon)= 9.6 \varepsilon^{-0.4}{\rm
keV}\cdot{\rm s}^{-1}\cdot{\rm sr}^{-1}\cdot{\keV}^{-1}$, the soft
excess around 1\keV is explained.  Note that it is not our primary
purpose here to find the best fit parameters because the single
power-law component is simply an extrapolation from the higher energy
band and also because the possible Galactic absorption is not taken
into account here.  Nevertheless it is interesting to see that the LG
halo can be a possible origin for the soft excess.

If the X-ray halo of the LG really exists and accounts for the soft
excess component in the XRB, it should also produce temperature
anisotropies in the cosmic microwave background (CMB) via the
Sunyaev-Zel'dovich (SZ) effect (Zel'dovich \& Sunyaev 1969; Cole \&
Kaiser 1988; Makino \& Suto 1993; Persi et al. 1995).  In the
Rayleigh-Jeans regime, the SZ temperature decrement is given by
\begin{equation}
\label{eq:dttsz}
\dttsz(\mu) = -2 {kT \over mc^2}\sigma_\T N_e(\mu),
\end{equation}
where $k$ is the Boltzmann constant, $m$ is the electron mass, $c$ is
the velocity of light, and $\sigma_T$ is the cross section of the
Thomson scattering. We compute the multipoles expanded in spherical
harmonics:
\begin{equation}
\label{eq:dttexpand}
\dtt(\theta,\varphi) = 
\sum_{l=0}^\infty \sum_{m=-l}^{l} a_l^m \, Y_l^m(\theta,\varphi).
\end{equation}
With eqs.(\ref{eq:ne}) and (\ref{eq:dttexpand}), one obtains
\begin{equation}
\label{eq:al0}
(a_l^0)_\sz = -2 \sqrt{(2l+1)\pi}{kT \over mc^2}\sigma_\T
\int_{-1}^{1} N_e(\mu) P_l(\mu)\, d\mu ,
\end{equation}
where $P_l(\mu)$ are the Legendre polynomials. Averaging over the sky,
the above SZ anisotropies are expected to contribute in quadrature to
the CMB anisotropies as
\begin{equation}
\langle \dtt^2 \rangle = { 1\over 4\pi} \sum_{l=0}^\infty
(2l+1) \langle |a_l^m|^2 \rangle 
= { 1\over 4\pi} \sum_{l=0}^\infty (a_l^0)_\sz^2 
\equiv \sum_{l=0}^\infty (T_{l,\sz})^2 .
\end{equation}
The corresponding monopole, dipole and quadrupole anisotropies reduce
to
\begin{eqnarray}
\label{eq:t0}
T_{0,\sz} &=& \pi \theta_c 
     \,\sigma_\T {kT \over mc^2} {n_0 r_c^2 \over x_0} ,\\
\label{eq:t1}
T_{1,\sz} &=& 2\sqrt{3} \left( 1 - {r_c \over x_0} \theta_c\right) 
     \,\sigma_\T {kT \over mc^2} {n_0 r_c^2 \over x_0} ,\\
\label{eq:t2}
T_{2,\sz} &=& {\sqrt{5}\pi \over 4}
\left[\theta_c - 3{r_c \over x_0} + 3\left({r_c \over x_0}\right)^2
\theta_c \right] 
     \,\sigma_\T {kT \over mc^2} {n_0 r_c^2 \over x_0} ,
\end{eqnarray}
where $\theta_c \equiv \tan^{-1}(x_0/r_c)$.  

The {\sl COBE} FIRAS data (Mather et al. 1994) imply that the Compton
y-parameter, $y$, should be less than $2.5\times10^{-5}$ (95\%
confidence level).  With eq.(\ref{eq:t0}), this upper limit is
translated to
\begin{equation}
\label{eq:n0limit}
n_0 r_c^2/x_0 < 1.1\times10^{22} 
 \left({1.17 \over \theta_c}\right)
 \left({y \over 2.5\times10^{-5}}\right)
 \left({1 \keV \over T}\right) \, {\rm cm}^{-2}.
\end{equation}
For example taking $r_c=0.15\mpc$ and $x_0=0.35\mpc$, the constraint
(\ref{eq:n0limit}) indicates that
\begin{eqnarray}
\label{eq:t1limit}
T_{1,\sz} &<& 3\times10^{-5} \left({y \over 2.5\times10^{-5}}\right) ,\\
\label{eq:t2limit}
T_{2,\sz} &<& 1.3\times10^{-5} \left({y \over 2.5\times10^{-5}}\right) .
\end{eqnarray}
The analysis of the 1st 2 years' {\sl COBE} DMR data (G\'orski 1994;
G\'orski et al. 1994; Bennett et al. 1994; Wright et al. 1994), on the
other hand, yields $T_{1,\COBE} = (1.23\pm0.09)\times10^{-3}$, and
$T_{2,\COBE} = (2.2\pm1.1)\times10^{-6}$. Therefore the LG X-ray halo
can potentially have significant effect on the quadrupole of the CMB
anisotropies, while its effect on dipole is totally negligible
compared to the peculiar velocity of the LG with respect to the CMB
rest frame.

\bigskip
\begin{center}
\section{\normalsize\bf IMPLICATIONS}
\end{center}
\setcounter{equation}{0}

Primordial density fluctuations with power spectrum $P(k) \propto k^n$
induce CMB anisotropies via the Sachs-Wolfe effect with multipoles
(e.g., Peebles 1993)
\begin{equation}
\label{eq:swcl}
C_l \equiv \langle |a_l^m|^2 \rangle \propto
{\Gamma(l+n/2-1/2) \over \Gamma(l-n/2+5/2)} ,
\end{equation}
where $\Gamma(\nu)$ is the Gamma function.  Therefore the
standard Harrison-Zel'dovich ($n=1$) spectrum predicts that
\begin{equation}
{l(l+1)C_l \over 6C_2} = 1 .
\end{equation}
The analytic expressions for the higher multipoles (eq.[\ref{eq:al0}])
are quite complicated. Instead we have numerically computed $(C_l)_\sz
\equiv (a_l^0)_\sz^2 /(2l+1)$ which are plotted in Fig.\ref{fig:szcl}. 
Clearly the higher multipoles decrease very rapidly with $l$. Thus
even if the CMB quadrupole were contaminated by the SZ effect
described here, the higher moments would be relatively free from such
an effect and can be interpreted to reflect the true cosmological
signature (the octapole may be affected to some extent). Although the
contribution of a distant cluster of galaxies to the multipoles is
small, its cumulative effect over the high-redshift may be observable
in the small-scale CMB anisotropies (Bennett et al. 1993; Makino \&
Suto 1993; Persi et al. 1995).

In this context it is interesting to note that the {\it rms}
quadrupole amplitude from the {\sl COBE} 2 years' data, $Q_{\rm rms} =
(6\pm3)\mu$K, is significantly smaller than that expected from the
higher multipoles (G\'orski 1994; G\'orski et al. 1994; Bennett et al. 
1994; Wright et al. 1994); if one fixes $n=1$, for instance, the power
spectrum fitting using eq.(\ref{eq:swcl}) requires that the most
likely amplitude should be $Q_{\rm rms-PS} = (18.2\pm1.5) \mu$K.  It
is somewhat common to ascribe the difference to cosmic variance. It is
possible, however, to account for it in terms of our model described
here, depending on the actual pattern of the primordial temperature
fluctuations.

In turn we can constrain the properties of the possible LG halo from
the COBE data. This is summarized in Fig. \ref{fig:constr}.  In fact,
the parameter range which is required for the LG halo to provide the
excess soft component is largely consistent with the current COBE
data. In addition, the LG X-ray halo should produce a dipole signature
(toward M31 and the opposite direction) in the soft excess component;
the flux $f$ plotted in Fig. 1 corresponds to what should be observed
at the center of the halo. If we adopt $x_0=0.35$Mpc, the flux towards
M31 should be $1.27f$ while $0.73f$ for the opposite direction.  Such
a level of XRB variation is detectable with careful data analysis of,
for instance, the ASCA GIS observation.  It should be noted that the
cooling time of the X-ray halo at the center due to the thermal
bremsstrahlung emission is roughtly estimated as
\begin{equation}
t_{cool} \equiv {3 n_0 kT \over \varepsilon_{ff}} 
\sim 3\times 10^{11} 
\left( {10^{-4} {\rm cm}^{-3} \over n_0} \right)
\sqrt {T \over 1 \keV} ~{\rm years} ,
\end{equation}
where $\varepsilon_{ff}$ is the thermal bremsstrahlung emission rate
per unit volume for the primordial gas.  Since $t_{cool}$ is
significantly larger than the age of the universe for the parameters
which we are interested in here, such a LG halo is expected to survive
for long once it is formed.

Our model described above assumes a fairly idealistic density profile
(\ref{eq:ner}). A more realistic profile of the halo including
non-sphericity and spatial inhomogeneity in temperature and density
will have a stronger effect on the higher multipoles
($l\geq3$). Therefore one might even probe the properties of the LG
halo through the multipoles of the CMB map.  The direct X-ray
detection of, or constraints on, the LG halo component is of great
importance in deriving the primordial spectral index $n$ and the
amplitude of the density fluctuations from the {\sl COBE} data.  It is
also important to search for the signature of, and/or put constraints
on, the LG halo from the direct analysis of the ROSAT and COBE data.

\bigskip
\bigskip

We are grateful to the referee, Renyue Cen, for his pertinent comments
on the earlier manuscript of the present {\it Letter}.  We also thank
Naoshi Sugiyama for discussions, and Ewan Stewart for a careful
reading of the manuscript. This research was supported in part by the
Grants-in-Aid by the Ministry of Education, Science and Culture of
Japan (07740183, 07CE2002).

\newpage

\baselineskip=24pt
\parskip2pt
\bigskip
\centerline{\bf REFERENCES}
\bigskip

\def\apjpap#1;#2;#3;#4; {\pp#1, {#2}, {#3}, #4}
\def\apjbook#1;#2;#3;#4; {\pp#1, {#2} (#3: #4)}
\def\apjppt#1;#2; {\pp#1, #2.}
\def\apjproc#1;#2;#3;#4;#5;#6; {\pp#1, {#2} #3, (#4: #5), #6}

\apjpap Bennett,C.L. et al. 1993;ApJL;414;L77;
\apjpap Bennett,C.L. et al. 1994;ApJ;436;423;
\apjpap Cen,R., Kang,H., Ostriker,J.P., \& Ryu,D. 1995;ApJ;451;436;
\apjpap Cole,S. \& Kaiser,N. 1988;MNRAS;233;637;
\apjpap Fabian,A.C. \& Barcons,X. 1992;ARA \& A;30;543;
\apjpap Gendreau,K.C. et al 1995;Pub.Astron.Soc.Japan.;47;L5;
\apjpap G\'orski 1994;ApJL;430;L85;
\apjpap G\'orski et al. 1994;ApJL;430;L89;
\apjproc Hasinger, G. 1992; The X-ray Background;
eds. Barcons, X. \& Fabian, A.C.;Cambridge University Press;
Cambridge;229;
\apjppt Kitayama,T. \& Suto,Y. 1996;MNRAS, in press;
\apjpap Makino,N. \& Suto,Y. 1993;ApJ;405;1;
\apjpap Mather,J.C. et al. 1994;ApJ;420;439;
\apjpap McCammon,D. \& Sanders, W.T. 1990;ARA \& A;28;657;
\apjbook Peebles,P.J.E. 1993;Principles of Physical Cosmology;
Princeton University Press;Princeton;
\apjpap Persi,F.M., Spergel,D.N., Cen,R., \& Ostriker,J.P. 1995;ApJ;442;1;
\apjpap Shanks, T. et al. 1991;Nature;353;315;
\apjpap Wang,Q.D., \& MaCray,R. 1993;ApJ;409;L37;
\apjpap Wright,E.L., Smoot,G.F., Bennett,C.L., \&
Lubin,P.M. 1994;ApJ;436;443;
\apjpap Wu,X., Hamilton,T, Helfand,D.J., \& Wang, Q. 1991;ApJ;379;564;
\apjpap Zel'dovich,Ya.B. \& Sunyaev,R.A. 1969;Ap.Sp.Sci.;4;301;

\newpage

\centerline{\bf FIGURE CAPTIONS}

\pp {\bf Figure 1 :} XRB spectra observed with different X-ray
satellites. The size of the symbols denote the error bars for ASCA
(Gendreau et al. 1995; crosses) and Einstein (Wu et al. 1991;
diamonds). The Rosat data (Hasinger 1992) are indicated by the region
between the two lines in the upper left.  The histogram in the lower
left indicates the emission from the X-ray halo of the Local Group
with $T=1\keV$, $r_c=0.15\mpc$, $n_0 = 10^{-4} {\rm cm}^{-3}$ and 0.3
times solar abundances observed at the center of the LG (computed with
the Raymond-Smith model).  This LG halo model has a total X-ray
luminosity of $10^{41} {\rm erg}\cdot{\rm s}^{-1}$ between $0.5 \keV$
and $4\keV$, and an electron column density of $6\times10^{20} {\rm
cm}^{-2}$.  The single power-law component extrapolated from the
higher energy band, $I(\varepsilon)= 9.6 \varepsilon^{-0.4}{\rm
keV}\cdot{\rm s}^{-1}\cdot{\rm sr}^{-1}\cdot{\keV}^{-1}$, is plotted
in the dotted line, while the overall prediction (the Local Group
emission plus the power-law component) is plotted in the solid line.
The Galactic absorption is not taken into account.

\bigskip
\pp {\bf Figure 2 :}
Assumed geometry of the X-ray halo of the Local Group.

\bigskip
\pp {\bf Figure 3 :} Multipoles induced by the Sunyaev-Zel'dovich
effect of the Local Group halo. The primordial Harrison-Zel'dovich
spectrum prediction $l(l+1)C_l/ (6C_2)=1$ is plotted in dotted line.

\bigskip
\pp {\bf Figure 4 :} Constraints on the density and the size of the
halo of the Local Group from CMB anisotropies. Dotted curves
correspond to $y=2.5\times10^{-5}$ (thick), $2.5\times10^{-5}/2$,
$2.5\times10^{-5}/4$, $2.5\times10^{-5}/8$, and $2.5\times10^{-5}/16$.
Solid curves correspond to $Q_{\rm sz}=9\mu$K , $6\mu$K (thick) ,
$3\mu$K , $1.5\mu$K , and $0.75\mu$K.

\newpage
\bigskip
\begin{figure}
\begin{center}
   \leavevmode\psfig{figure=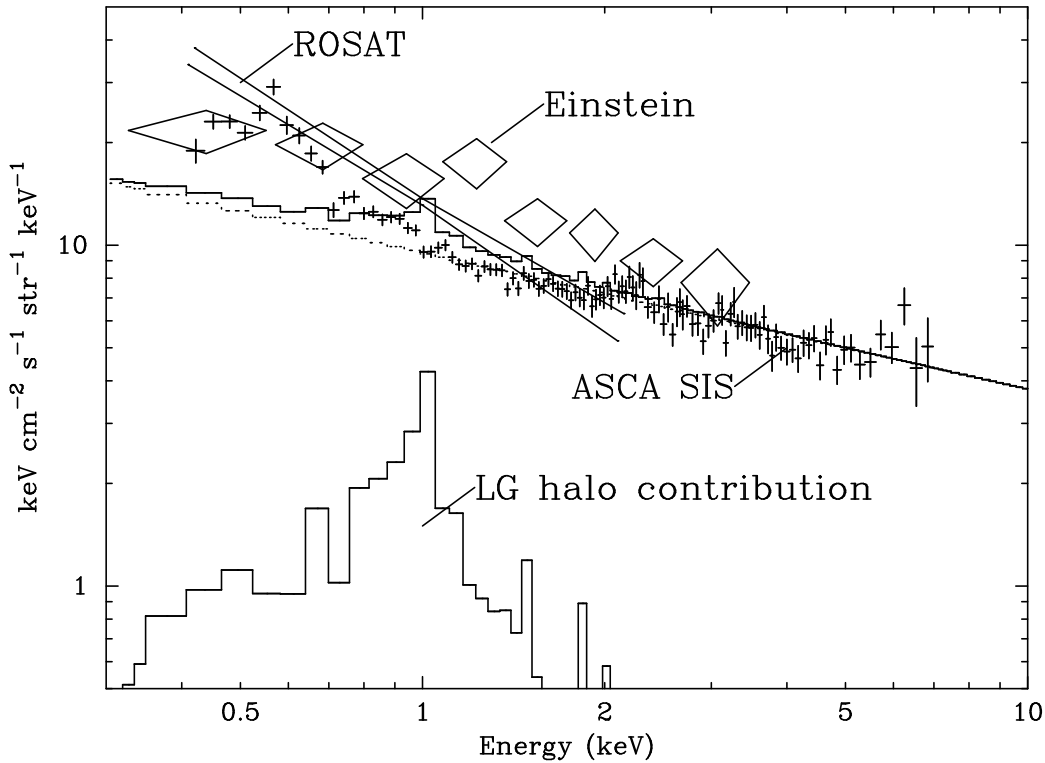,height=10cm,angle=-90}
\end{center}
\caption{XRB spectra observed with different X-ray satellites. The
size of the symbols denote the error bars for
ASCA (Gendreau et al. 1995; crosses) and
Einstein (Wu et al. 1991; diamonds). The Rosat
data (Hasinger 1992) are indicated by the region
between the two lines in the upper left.  The histogram in the lower
left indicates the emission from the X-ray halo of the Local Group
with $T=1\keV$, $r_c=0.15\mpc$, $n_0 = 10^{-4} {\rm cm}^{-3}$ and 0.3
times solar abundances observed at the center of the LG (computed with
the Raymond-Smith model).  This LG halo model has a total X-ray
luminosity of $10^{41} {\rm erg}\cdot{\rm s}^{-1}$ between $0.5 \keV$
and $4\keV$, and an electron column density of $6\times10^{20} {\rm
cm}^{-2}$.  The single power-law component extrapolated from the
higher energy band, $I(\varepsilon)= 9.6 \varepsilon^{-0.4}{\rm
keV}\cdot{\rm s}^{-1}\cdot{\rm sr}^{-1}\cdot{\keV}^{-1}$, is plotted
in the dotted line, while the overall prediction (the Local Group
emission plus the power-law component) is plotted in the solid line.
The Galactic absorption is not taken into account.  }
\end{figure}

\newpage
\bigskip
\begin{figure}
\begin{center}
   \leavevmode\psfig{figure=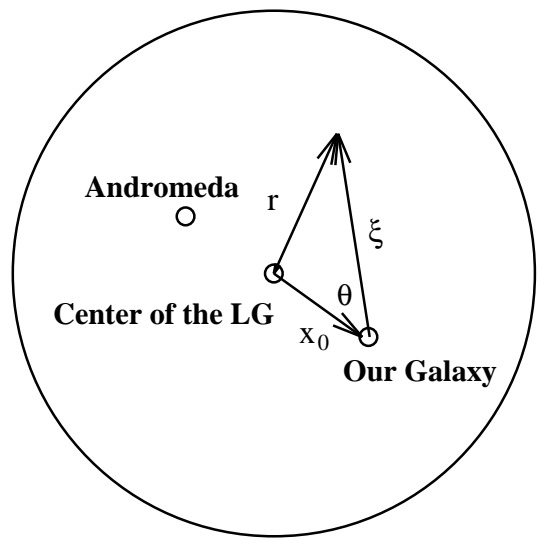,height=7cm,angle=-90}
\end{center}
\caption{Assumed geometry of the X-ray halo of the Local Group.
\label{fig:lghalo}}
\end{figure}

\newpage
\bigskip
\begin{figure}
\begin{center}
   \leavevmode\psfig{figure=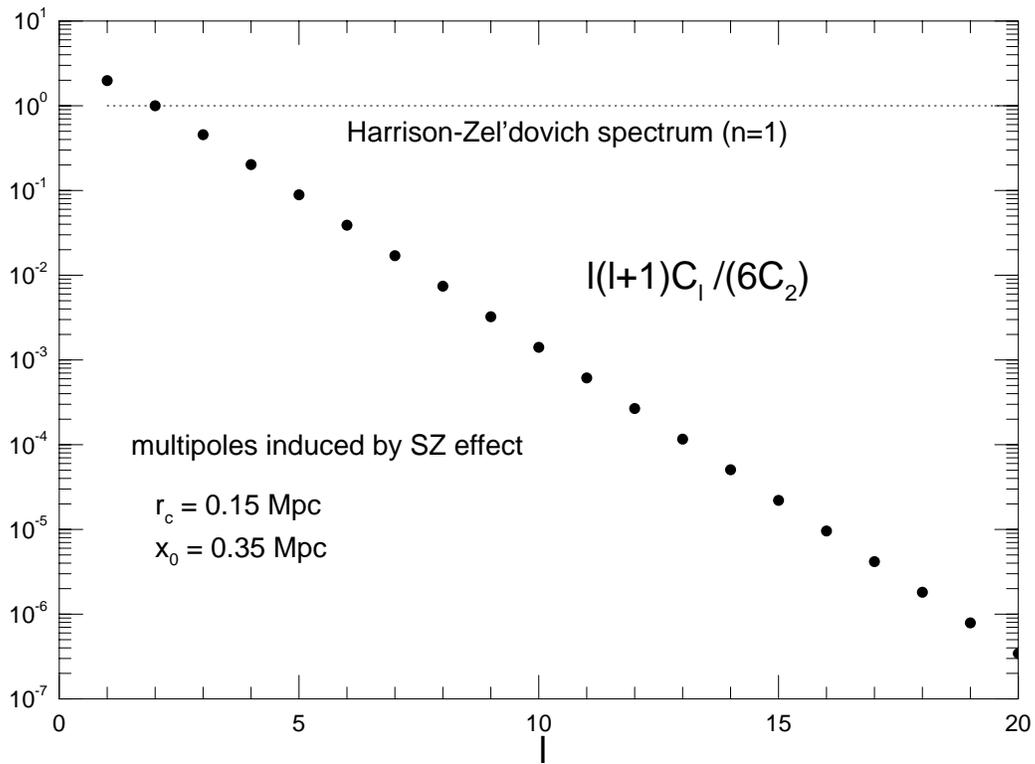,height=10cm,angle=90}
\end{center}
\caption{ Multipoles induced by the Sunyaev-Zel'dovich effect of the
Local Group halo. The primordial Harrison-Zel'dovich spectrum
prediction $l(l+1)C_l/ (6C_2)=1$ is plotted in dotted line.
\label{fig:szcl}
 }
\end{figure}

\newpage
\bigskip
\begin{figure}
\begin{center}
   \leavevmode\psfig{figure=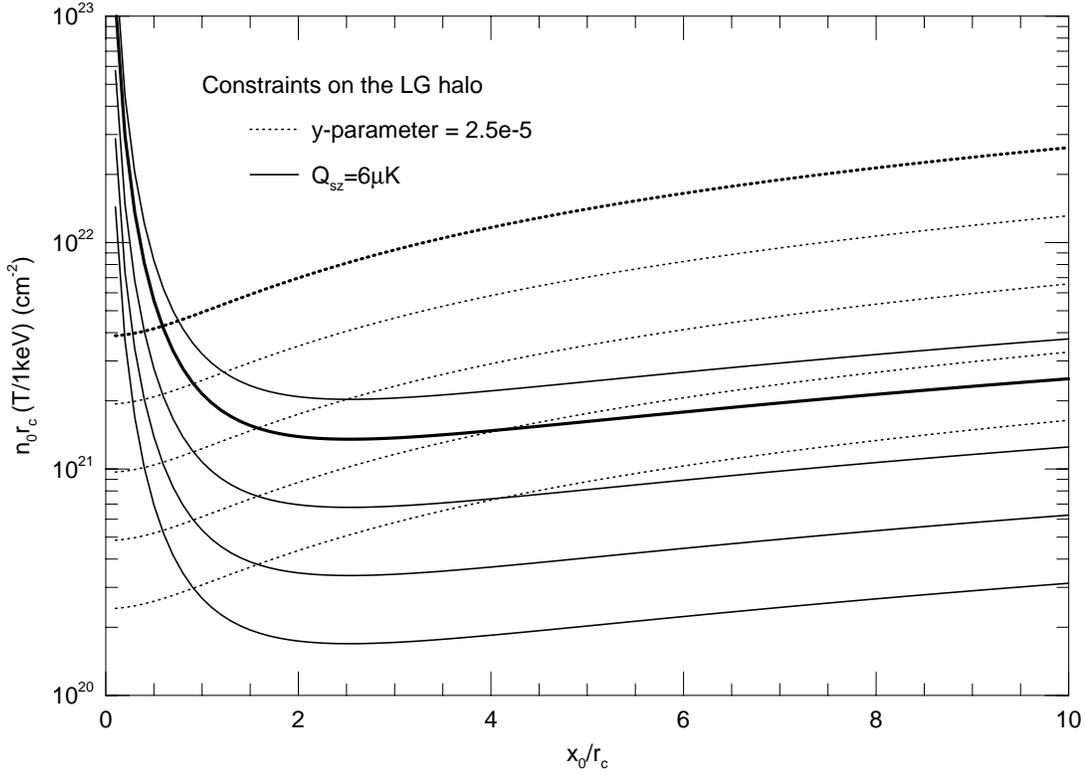,height=10cm,angle=180}
\end{center}
\caption{ Constraints on the density and the size of the halo of the
Local Group from CMB anisotropies. Dotted curves correspond to
$y=2.5\times10^{-5}$ (thick), $2.5\times10^{-5}/2$,
$2.5\times10^{-5}/4$, $2.5\times10^{-5}/8$, and $2.5\times10^{-5}/16$.
Solid curves correspond to $Q_{\rm sz}=9\mu$K , $6\mu$K (thick) ,
$3\mu$K , $1.5\mu$K , and $0.75\mu$K.
\label{fig:constr}
 }
\end{figure}

\end{document}